\documentclass[11pt]{article}
\usepackage{amssymb,amsmath,graphicx}
\usepackage[latin1]{inputenc}
\usepackage[english]{babel}  
\usepackage{float}
\usepackage{tikz}
\usepackage{dcolumn}
\usepackage{multirow}
\usepackage[round,authoryear]{natbib}
\usepackage{color, colortbl}
\usepackage{multirow}
\usepackage{microtype}
\usepackage{lscape}
\usepackage{enumitem}
\usepackage[ruled,linesnumbered,noend]{algorithm2e}
\usepackage{setspace}
\SetArgSty{textrm} 
\DontPrintSemicolon  
\usepackage[hidelinks]{hyperref}
\usepackage{siunitx} 
 \bibpunct[, ]{(}{)}{,}{a}{}{,}%
 \def\newblock{\ }%

\newcolumntype{d}[1]{D{.}{.}{#1}}

\newtheorem{proposition}{Proposition}

\newcommand{\cO}{{\mathcal O}}
\newcolumntype{M}[1]{>{\centering}m{#1}}
\newcommand{\cA}{{\cal A}}
\newcommand{\cG}{{\cal G}}
\newcommand{\cV}{{\mathcal V}}

\newcolumntype{P}[1]{>{\centering}p{#1}}
\newcolumntype{M}[1]{>{\centering}m{#1}}
\newcolumntype{K}[1]{>{\flushright}m{#1}}
\definecolor{LightGray}{gray}{0.92}

\newcommand{\blue}[1]{{#1}}

\newcommand{\setmuskip}[2]{#1=#2\relax}

\title{Technical Note: Split algorithm in O(n) for the capacitated vehicle routing problem}

\author{Thibaut Vidal}

\begin{document}

\vspace*{-1.8cm}

\begin{scriptsize}
\noindent
This is the post-peer-review, pre-copyedit version of the article published in \emph{Computers \& Operations Research}, 69, 40--47 (2016). The published article is available at \url{http://dx.doi.org/10.1016/j.cor.2015.11.012}. This manuscript version is made available under the CC-BY-NC-ND 4.0 license \url{http://creativecommons.org/licenses/by-nc-nd/4.0/}.\par
\end{scriptsize}

\vspace*{0.8cm}

\begin{center}

\begin{LARGE}
{\textbf{Technical Note: Split algorithm in O(n) \\ \vspace*{0.2cm}  for the \blue{capacitated} vehicle routing  problem}}
\end{LARGE}

\vspace*{1.2cm}

\textbf{Thibaut Vidal} \\
Departamento de Inform\'atica, Pontif\'icia Universidade Cat\'olica do Rio de Janeiro (PUC-Rio) \\
Rua Marqu\^es de S\~ao Vicente, 225 - G\'avea, Rio de Janeiro - RJ, 22451-900, Brazil \\
vidalt@inf.puc-rio.br \\

\vspace*{1cm}

\end{center}
\noindent
\textbf{Abstract.}
The Split algorithm is an essential building block of route-first cluster-second heuristics and modern genetic algorithms for vehicle routing problems. \blue{The algorithm is used to partition a solution, represented as a giant tour without occurrences of the depot, into separate routes with minimum cost.}
As highlighted by the recent survey of [Prins, Lacomme and Prodhon, Transport Res. C (40), 179--200], no less than 70 recent articles use this technique. In the vehicle routing literature, Split is usually assimilated to the search for a shortest path in a directed acyclic graph $\mathcal{G}$ and solved in $\cO(nB)$ using Bellman's algorithm, where $n$ is the number of delivery points and \blue{$B$ is the average number of feasible routes that start with a given customer in the giant tour.}
Some linear-time algorithms are also known for this problem as a consequence of a Monge property of~$\mathcal{G}$. In this article, we highlight a stronger property of this graph, leading to a \blue{simple} alternative algorithm in $\cO(n)$. Experimentally, we observe that the approach is faster than the classical Split for problem instances of practical size. We also extend the method to deal with a limited fleet and soft capacity constraints.
\vspace*{0.3cm}


\vspace*{0.3cm}

\section{Introduction}

The algorithm of \cite{Prins2004} was an important milestone for the vehicle routing problem (VRP): it was the first hybrid genetic algorithm with local search to outperform classical tabu searches at a time when such methods were predominant.
One main ingredient of its success was its approach to solution representation and recombination. 
Until the 2000s, combining two solutions was considered a difficult task, because simple crossover operators had a tendency to produce infeasible and unbalanced routes. 
To meet this challenge, \cite{Prins2004} represented the solution as a permutation of visits, a ``giant tour'', and relied on a dynamic-programming-based decoder, called \emph{Split}, which optimally inserts depot visits to obtain complete solutions.
This makes it possible to efficient use classical crossovers for permutations, since the Split algorithm is in charge of route delimitations, and the capacity constraints are implicitly managed during solution~decoding.

\blue{
Ten years on, the literature on population-based methods for VRPs has grown extensively.
Efficient GAs with a complete solution representation and more advanced crossover operators now exist for the capacitated VRP (e.g.,  \citealt{Nagata2009}), a sign that the Split algorithm is useful but not a necessity. Nevertheless, the approach of \cite{Prins2004} remains simple and generic. The representation as a giant tour enables to significantly reduce the number of distinct individuals in the GA, and many side constraints and auxiliary decisions of VRP variants, such as capacity and duration limits, time windows \citep{Vidal2012c}, choices of depots \citep{Duhamel2010}, vehicle types \citep{Duhamel2011a}, or profitable customers in each route \citep{Vidal2014} can be handled in the Split algorithm rather than in the crossover.
As such, Split has led to successful heuristics for a large number of problems, as surveyed in \cite{Duhamel2011a,Vidal2012,Prins2014}, and \cite{Laporte2014a}.
}

\blue{The computational efficiency of the Split algorithm for the Capacitated VRP (CVRP) is the subject of  this article. The CVRP aims to find minimum-distance routes to service $n$ customer locations with respective demands $q_1,\dots,q_n$, using a fleet of up to $m$ vehicles of capacity $Q$ located at a central depot. Here, we consider that an input solution is given, represented as a giant tour $(1,\dots,n)$ (w.l.o.g., the visits are re-indexed by order in the tour). Let $d_{i,i+1}$ be the distance between two successive customers, and $d_{0i}$ and $d_{i0}$ be the distances from and to the depot. All distances and demand quantities are assumed to be non-negative. The objective of Split is to partition the giant tour into $m$ disjoint sequences of consecutive visits. Each such sequence is associated to a route, which originates from the depot, visits its respective customers, and returns to the depot. The total distance of all routes should be minimized. Note that the algorithms of this paper do not require the symmetry of the distance matrix or the triangle inequality.}

\blue{Classically, the Split algorithm is reduced to a shortest path problem between the nodes $0$~and~$n$ of an acyclic graph $\cG = (\cV,\cA)$, where $\cV = (0,\dots,n)$, and $\cA$ contains one arc $(i,j)$ with cost $c(i,j) = d_{0,i+1} + \sum_{k=i+1,\dots,j-1} d_{k,k+1} + d_{j,0}$
for any feasible route visiting customers $i+1$~to~$j$.
In the literature, the shortest path is obtained in $\cO(nB)$ via a variant of Bellman's algorithm, where $B$ is the average out-degree of a node in $\{0,\dots,n-1\}$, i.e., the average number of feasible trips from one node of the giant tour \citep{Beasley1983,Prins2004}.}
Moreover, for a limited fleet of~$m$ vehicles, the propagation of the labels can be iterated to produce a shortest path with at most~$m$ arcs in $\cO(nmB)$. Such complexity is suitable for most medium-scale applications. However, Split can become a computational bottleneck for large problems \blue{with many deliveries per route}, when used iteratively in a metaheuristic.

To meet this challenge, we will introduce a new \emph{Split} algorithm in $\cO(n)$.
Note that some linear-time algorithms are already known for this shortest path \citep[see][and the references therein]{Burkard1996,Bein2005} as the graph $\cG$ satisfies the Monge property:
\begin{equation}
\begin{aligned}
&& c(i_1,j_1) + c(i_2,j_2) \leq c(i_1,j_2) + c(i_2,j_1)  &\text{ for all } 0 \leq i_1 < i_2 < j_1 < j_2 \leq n  \\
&&&\blue{\text{ such that } (i_1,j_2) \in \cA},
\end{aligned}
\end{equation}
where $c(i,j)$ is the cost of an arc $(i,j)$.
So far, these methods were not applied~in~the~VRP~literature.

In this article, we propose a simpler alternative which uses the fact that the auxiliary graph $\cG$ satisfies the following stronger property:
\begin{equation}
\begin{aligned}
&\text{for all } 0 \leq i_1 < i_2 < n, \text{ there exists } K  \in \mathbb{R} \text{ such that } \\
 &c(i_1,j) - c(i_2,j) = K \   \text{for all }   j  > i_2 \text{ such that } (i_1,j) \in \cA.
\end{aligned} \label{eqequa}
\end{equation}
We show that Property (2) can be used to eliminate dominated predecessors and retain only good candidates, leading to a very simple labeling algorithm in $\cO(n)$ which performs well in practice and can be efficiently used in VRP metaheuristics.
The approach is also extended to produce a solution of the Split problem with a limited number of vehicles in $\cO(nm)$, and with soft capacity constraints~in~$\cO(n)$. 

Finally, we compare the practical CPU time of the proposed method with that of the classical Bellman-based algorithm, using giant tours built from TSP instances. These instances contain from $n=29$ to 71,009 nodes, and the number of deliveries per route ranges from $4$ to $4,000$.  The linear approach appears to be faster in most cases, with speedup factors ranging from $0.8$ to $400$. The largest speedups are achieved for instances with many deliveries per route, which can occur in courier delivery, refuse collection, and meter reading applications.

The remainder of this paper recalls the Bellman-based Split algorithm in Section~2, introduces the proposed linear Split in Section~3, discusses its generalization to limited fleets and soft capacity constraints in Section~4, and reports our computational experiments in Section~5. To facilitate the use of these algorithms in future generations of heuristics, a C++ implementation of the methods of this paper is available at \url{http://w1.cirrelt.ca/\~vidalt/en/VRP-resources.html}. 

\section{Bellman-based Split Algorithm}
\label{sectionBellman}

Split is traditionally based on a simple dynamic programming algorithm, which enumerates the nodes in topological order and, for each node $t$, propagates its label to all successors $i$ such that $(t,i) \in \cA$. The presentation in Algorithm~\ref{algoSimple} is similar to that of \cite{Prins2004}. The arc costs are not preprocessed but directly computed in the inner loop. This specific algorithm was used as a benchmark in our computational experiments in Section \ref{sectionExperiments}.

\begin{algorithm}[htbp]
\setstretch{1.05}
$p[0] \gets 0$ \;
\For{$t = 1$ to $n$}
{
	$p[t] \gets \infty$ \;
}
\For{$t = 0$ to $n-1$}
{ 
	$load \gets 0$ \;
	$i \gets t+1$ \;
	\While{$i \leq n$ \textbf{and} $load + q_i \leq Q$}
	{ 
		$load \gets load + q_i$ \;
		\If{$i = t+1$}
		{
			$cost \gets d_{0,i}$ \;
		}
		\Else
		{
			$cost \gets cost + d_{i-1,i}$ \;
		}
     	  	 \If{$p[t] + cost + d_{i0} < p[i]$}
		{
			$p[i] = p[t] + cost + d_{i0}$ \;
            	        $pred[i] = t$ \;
		}
		$i \gets i+1$ \;
	}
}
\caption{Classical Split Algorithm} 
\label{algoSimple} 
\end{algorithm}

At the end of each iteration $t$ (lines 5--16 of Algorithm~\ref{algoSimple}), $p[t]$ contains the cost of a shortest path from~$0$~to~$t$. The array of predecessors $pred$ is maintained throughout the search so that we can retrieve the solution at the end of the algorithm.

\section{Split in Linear Time}
\label{sectionLinear}

This section will introduce a more efficient Split algorithm.
\blue{As in the classical Split, the arc costs of the underlying graph are not pre-processed.
We will describe, in turn, some auxiliary data structures, the data for a numerical example, and the proposed algorithm}. \\

\noindent
\textbf{Preliminaries.} We define for $i \in \{1,\dots,n\}$ the cumulative distance $D[i]$ and cumulative load $Q[i]$ as follows:
\begin{align}
D[i] &=  \sum_{k=1}^{i-1} d_{k,k+1} \label{cumulativeDistance} \\
Q[i] &= \sum_{k=1}^{i} q_k.   \label{cumulativeLoad}
\end{align}

These values can be preprocessed and stored in $\cO(n)$ at the beginning of the algorithm.
For $i < j$, the cost $c(i,j)$ of an arc $(i,j)$ is the cost of leaving the depot, visiting customers $(i+1,\dots,j)$, and returning to the depot, computed as
\begin{equation}
c(i,j) = d_{0,i+1} + D[j] - D[i+1] + d_{j,0},
\end{equation}
and the arc $(i,j)$ exists if and only if the route is feasible, i.e., $Q[j]-Q[i] \leq Q$. \\

Our algorithm also relies on a double-ended queue, denoted $\Lambda$, that supports the following operations in $\cO(1)$:

\noindent
\hspace*{1cm}
\begin{tabular}{r@{ -- }l}
\emph{front} &accesses the oldest element in the queue; \\
\emph{front2} &accesses the second-oldest element in the queue; \\
\emph{back} &accesses the most recent element in the queue; \\
\emph{push\_back} &adds an element to the queue; \\
\emph{pop\_front} &removes the oldest element in the queue; \\
\emph{pop\_back} &removes the newest element in the queue. \\
\end{tabular}

\noindent
We will refer to the elements of the queue as $(\lambda_1,\dots,\lambda_{|\Lambda|})$, from the front $\lambda_1$ to the back $\lambda_{|\Lambda|}$. \\

\begin{figure}[H]
\begin{minipage}[t]{.65\linewidth}
\renewcommand{\arraystretch}{1.12}
\setlength{\tabcolsep}{0.12cm}
\scalebox{0.9}
{
\begin{tabular}{|c|ccccccccccccc|}
\hline
\textbf{Node}&\textbf{0}&\textbf{1}&\textbf{2}&\textbf{3}&\textbf{4}&\textbf{5}&\textbf{6}&\textbf{7}&\textbf{8}&\textbf{9}&\textbf{10}&\textbf{11}&\textbf{12}\\
\hline
$d_{i-1,i}$&---&4&3&7&2&7&3&8&6&8&4&3&3\\
$d_{0,i}$&---&4&5&10&9&14&12&16&11&5&3&5&6\\
$q_i$&---&11&3&6&5&7&8&1&7&3&7&3&6\\
\hline
\hline
$p[i]$&\textbf{0}&\textbf{8}&\textbf{12}&\textbf{24}&\textbf{25}&\textbf{43}&\textbf{44}& 56 & 67& 69& 75 & 80 & 84 \\
\hline
\end{tabular}
}
\end{minipage}
\begin{minipage}[c]{.35\linewidth}
\vspace*{-0.5cm}
\includegraphics[width=5.2cm]{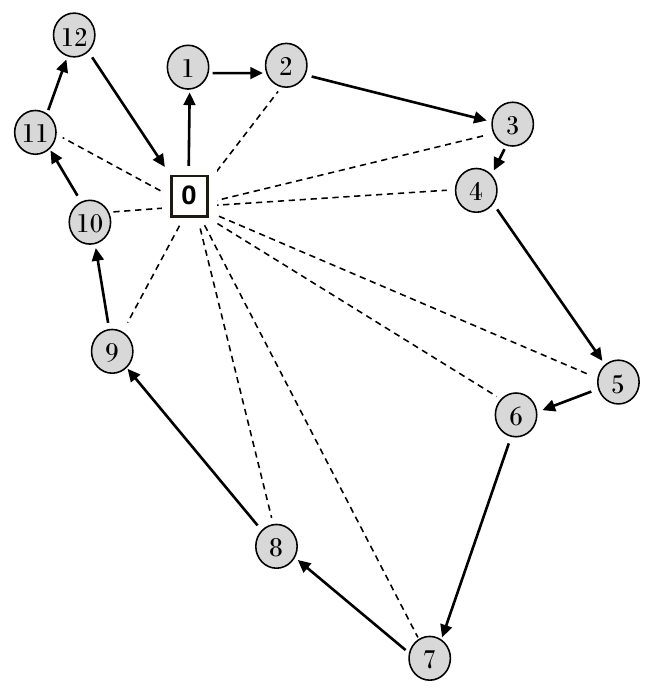}
\vspace*{-1cm}
\end{minipage}
\caption{Input data for the Split algorithm and values of $p[i]$.\hspace*{5cm}}
\label{figure-numericalExample}
\end{figure}

\noindent
\textbf{Data for the Numerical Example.}
To illustrate the algorithm, we use a numerical example with 12 nodes.
Figure \ref{figure-numericalExample} provides the input distances from the depot and between successive nodes as well as the demands associated with each node.  For this instance, the best solution consists of the routes (0,1,2,3,4,0), (0,5,6,7,8,9,0), and (0,10,11,12,0).

\noindent
\textbf{Main algorithm.}  \blue{Instead of iterating over all arcs to propagate minimum-cost paths, the proposed Algorithm \ref{algoLinearSplit} takes advantage of the cost structure of the Split graph and maintains a set of \emph{nondominated predecessors} in a queue $\Lambda$ (lines 7--12). For each nodes $t \in \{1,\dots,n\}$, this structure enables to find in~$\cO(1)$ a best predecessor for $t$ along with the cost of a shortest path from $0$ to $t$ (line 4).}

\begin{algorithm}[htb]
\setstretch{1.1}

$p[0] \gets 0$ \;
$\Lambda \gets  (0)$ \;
\For{$t = 1$ to $n$}
{ 

  	$p[t] \gets p[front] + f(front,t)$ \;   \label{linearSplitPropagate}
  	$pred[t] \gets front $ \;

 	 \If{$t < n$}
	{
		 \If{\textbf{not} $dominates(back,t)$}
		{
			\While{$|\Lambda| > 0$  \textbf{and}  $dominates(t,back)$ }
	  		{ $popBack()$ \; }
			$pushBack(t)$
		}

   		\While{$Q[t+1] > Q + Q[front]$}
  		{ $popFront()$ \; }
	}
}
 \caption{Linear Split} 
\label{algoLinearSplit} 
\end{algorithm}

The definition of the boolean function $dominates(i,j)$ completes the algorithm. \blue{This function returns \textsc{True} if and only if the node $i$ dominates the node $j$ as a predecessor.}
\begin{equation}
\begin{small}
dominates(i,j) \equiv
\begin{cases}
p[i] +d_{0,i+1} - D[i+1] \leq p[j] + d_{0,j+1} - D[j+1]   \textbf{ and } Q[i] = Q[j]  & \text{ if } i \leq j \\
p[i] + d_{0,i+1} - D[i+1] \leq  p[j] + d_{0,j+1} - D[j+1] & \text{ if } i > j. \\
\end{cases} 
\end{small}
\label{dominatesFunctionHard}
\end{equation}

\noindent
\textbf{Correctness of the algorithm.}
Define $f(i,x)$ as the cost achieved when extending the label of a predecessor $i$ to a node $x \in \{i+1, \dots, n\}$. This function takes an infinite value if the arc $(i,x) \notin \cA$ because of capacity constraints:
\begin{equation*}
f(i,x) =
\begin{cases}
p[i] + c(i,x)  & \hspace*{0.3cm}  Q[x] - Q[i] \leq Q \\ 
\infty & \hspace*{0.3cm} otherwise.
\end{cases} 
\end{equation*}


Furthermore, define the auxiliary function $g_i(x) = f(i,x) - D[x] - d_{x0}$. This function of $x$ takes a constant value as long as the label extension is feasible, since
\begin{equation}
\begin{aligned}
\textbf{ if }  Q[x] - Q[i] \leq Q, \hspace{0.3cm} g_i(x) &= p[i] + d_{0,i+1} + D[x] - D[i+1] + d_{x0} - D(x) - d_{x0} \\
&= p[i] + d_{0,i+1} - D[i+1].
\end{aligned}
\end{equation}

\begin{figure}[htbp]
\centering
\includegraphics[width=7.5cm]{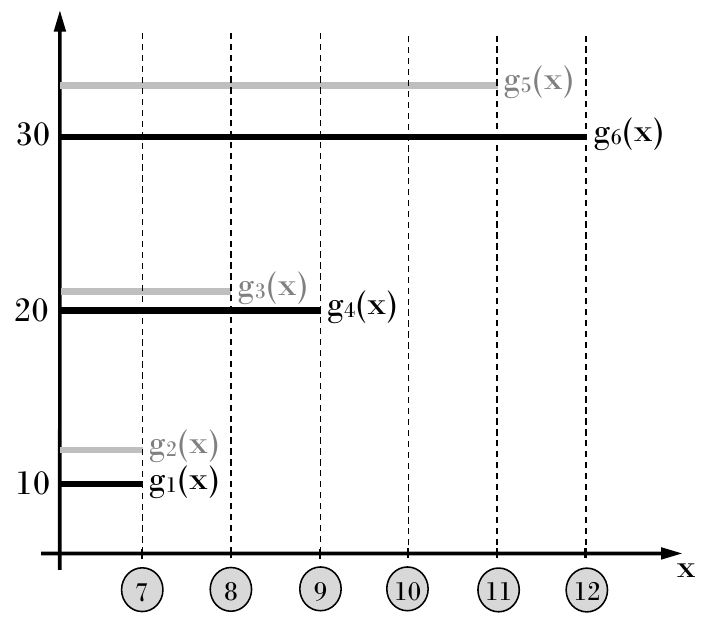}
\caption{Numerical example -- The functions $g_i(x)$ are displayed for $i \in \{1,\dots,6\}$. Note that the function $g_0(x)$ does not appear (takes an infinite value) since $(0,7) \notin \cA$.}
\label{figure-hard}
\end{figure}

This observation corresponds to the property announced in the introduction, in Equation (\ref{eqequa}).
Figure \ref{figure-hard} displays the functions $g_i(x)$ of the numerical example at iteration $t=7$.

Now, consider two candidate predecessors $i$ and $j$ such that $i < j < t$. The expression
$f(i,x) - f(j,x)$ represents the cost difference between extending the predecessor $i$ to a node $x \geq t$ and extending $j$ to the same node. Note that $f(i,x) - f(j,x) = g_i(x) - g_j(x)$. Therefore, each predecessor $i < t$ can be characterized by:
\begin{itemize}[nosep]
\item[]
\begin{itemize}[nosep]
\item a \textbf{fixed cost} $g_i = p[i] + d_{0,i+1} - D[i+1]$,
\item and the \textbf{cumulative demand} $l_i = Q[i]$.
\end{itemize}
\end{itemize}
These two values lead to the dominance relation of Equation (\ref{dominatesFunctionHard}): if $g_i \geq g_j$ and $l_i \geq l_j$, then $g_i(x) \geq g_j(x)$ for any $x \geq t$ and candidate $j$ is always at least as good as $i$.

Algorithm \ref{algoLinearSplit} has been designed to satisfy the following invariant at line~4:
\begin{proposition}[Loop Invariant]
$\Lambda$ contains mutually nondominated predecessors of $t$ ranked by increasing cost: for $i \in \{1,\dots,|\Lambda|-1\}$, $g_{\lambda_i} < g_{\lambda_i+1}$ and $l_{\lambda_i} \leq l_{\lambda_i+1}$.  Moreover, $Q[t] \leq Q + l_{\lambda_1}$.
\end{proposition}
This proposition is true for $t=1$ since $\Lambda = (0)$ and $q_1 \leq Q$. 
Then, any new node $t$ will be added to $\Lambda$ only if it is not dominated by the current back element (lines 7 and 10). If $t$ is added, the loop at lines 8--9 removes from the queue any node that is dominated by~$t$, starting from the back of the queue and stopping as soon as the first nondominated element is found. Since the elements are ordered by increasing cost, this guarantees the removal of all predecessors dominated by $t$. Finally, when the index $t$ is incremented, any predecessor at the front of the queue that cannot be feasibly extended to $t+1$ is eliminated to ensure that $Q[t+1] \leq Q + l_{\lambda_1}$.

As a consequence of this invariant, the first element of the queue is always a best predecessor for~$t$. Indeed, it is a feasible predecessor, and all other predecessors in the queue have a greater cost. Furthermore, any other element that is no longer in $\Lambda$ was either dominated or could not be feasibly extended to a node $x \geq t$. This proves the correctness of the algorithm.

In terms of complexity, we remark that each node $i$ cannot be added to the queue more than once via \emph{pushBack} or deleted more than once via $popBack$ or $popFront$. As a consequence, the operations of lines 7--12 are performed at most $n$ times, leading to an overall complexity~of~$\cO(n)$.

\section{Extensions of the Linear Split Algorithm}

This section describes two extensions of the proposed algorithm, for the Split problem in the presence of a limited number of vehicles, and for the case where linear penalties are imposed if the capacity is exceeded.

\subsection{Limited Number of Vehicles}
\label{sectionLinearLimited}

The extension of the algorithm to a limited number of vehicles requires us to perform the previous algorithm once for each vehicle. The resulting approach is described in Algorithm \ref{algoLinearSplit-Limited}.

\begin{algorithm}[htbp]
\setstretch{1.1}

\For{$k = 1$ to $m$}
{
\For{$t = 0$ to $n$}
{
$p[k,t] = \infty$ \;
}
}

$p[0,0] \gets 0$ \;
\For{$k = 0$ to $m-1$}
{ 
	$clear(\Lambda)$ \;
	$\Lambda \gets (k)$ \;
	\For{$t = k+1$ to $n$ \textbf{s.t.} $|\Lambda| > 0$}
	{ 
  		$p[k+1,t] \gets p[k,front] + f(front,t)$ \;   \label{linearSplitPropagate-Limited}
  		$pred[k+1][t] \gets front $ \;

 	 	\If{$t < n$}
		{
			 \If{\textbf{not} $dominates(k,back,t)$}
			{
				\While{$|\Lambda| > 0$  \textbf{and}  $dominates(k,t,back)$ }
	  			{ $popBack()$ \; }
				$pushBack(t)$
			}
   			\While{$|\Lambda| > 0$  \textbf{and} $Q[t+1] > Q + Q[front]$}
  			{ $popFront()$ \; }
		}
	}
}
 \caption{Linear Split: Fleet limited to $m$ vehicles} 
\label{algoLinearSplit-Limited} 
\end{algorithm}

For $k \in \{1,\dots m\}$ and $t \in \{k,\dots,n\}$, the two-dimensional array $p[k,t]$ will contain the cost of a shortest path with $k$ arcs finishing at $t$. These costs are computed for increasing $k$ in an outer loop (line 5) and for increasing $t$ in the inner loop (line 8). The cost of any label $p[k+1,t]$ is obtained from the extension of a best predecessor $p[k,i]$ with $i < t$. Applying the Bellman algorithm in the inner loop would lead to a complexity of $\cO(n^2 m)$. Instead, we use the queue data structure and dominance properties as in the previous section, leading to a complexity of $\cO(n)$ in the inner loop, for a total complexity of $\cO(nm)$.

The inner loop can be stopped once $\Lambda$ is empty (line 8). In this state, the algorithm has reached the last index that can be feasibly attained with $k$ routes.
The minimum cost of a route containing $m$ vehicles is given at the end of the algorithm by $p[m,n]$, and the two-dimensional array $pred$ enables us to trace back the solution. 
The minimum cost of a route containing $k \leq m$ vehicles can also be found, by seeking the minimum of $p[k,n]$, for $k \in \{1,\dots,m\}$.
Note that the \emph{dominates} function takes the number of vehicles $k$ as an extra argument---since it considers the two-dimensional array $p[k,i]$ instead of $p[i]$---but its purpose remains the same.

\subsection{Soft Capacity Constraints}
\label{sectionLinearSoft}

Consider the case where the capacity of a route may be exceeded, subject to a linear penalty with coefficient $\alpha \geq 0$. This relaxation of the capacity constraints is useful in practical situations where the \emph{demand} of a customer represents a time or workload quantity rather than a physical load in a truck, and where an excess may be acceptable. This relaxation is also useful in heuristics, allowing them to better explore the search space via intermediate infeasible solutions and adaptive penalties \citep{Gendreau1994,Cordeau1997,Vidal2013a}.

All arcs $(i,j)$ such that $i < j$ are now included in $\cA$, and the cost of an arc $(i,j)$ is
\begin{equation}
c(i,j) = d_{0,i+1} + D[j] - D[i+1] + d_{j,0} + \alpha \times \max \{ Q[j] - Q[i] - Q , 0 \}.
\end{equation}

\noindent
\textbf{Main algorithm.}
Algorithm \ref{algoLinearSplit} can still be applied subject to two changes:
\begin{enumerate}
\item The \emph{dominates} function is updated to account for the constraint relaxations:
\begin{equation}
\hspace*{-1.4cm}
\begin{small}
dominates(i,j) \equiv
\begin{cases}
p[i] + d_{0,i+1} - D[i+1] + \alpha \times (Q[j] - Q[i]) \leq p[j] + d_{0,j+1} - D[j+1]   & \text{ if } i < j \\
p[i] + d_{0,i+1}  - D[i+1] \leq  p[j] + d_{0,j+1} - D[j+1] & \text{ if } i > j. \\
\end{cases} 
\end{small}
\label{dominatesFunctionSoft}
\end{equation}
\item The rule for eliminating the front label in $\Lambda$, at line 11, becomes:
\end{enumerate}

\begin{center}
\textbf{while} $|\Lambda| > 1$  \textbf{and} $p[front] + f(front,t+1) \geq  p[front2] + f(front2,t+1)$. \\ \vspace*{0.3cm}
\end{center}

\noindent
\textbf{Correctness of the algorithm.}
We rely on the same principles as before.
The cost $f(i,x)$ of the extension of a node $i$ to a node $x \in \{i+1,\dots,n\}$ and the functions $g_i(x)$ are  defined as:
\begin{align}
f(i,x) &= p[i] + c(i,x) \\
g_i(x) &= f(i,x) - D[x] - d_{x0} \\
	&= p[i] + d_{0,i+1} - D[i+1] + \alpha \times \max \{ Q[x] - Q[i] - Q , 0 \}.
\end{align}

Again, $g_i(x) \leq g_i(x)$ means that $j$ is dominated by $i$ as a predecessor. 
Now, we define the function $h_i(y)$ for $y \in \mathbb{R}$ as:
\begin{equation}
h_i(y) = p[i] +  d_{0,i+1} - D[i+1] + \alpha \times \max \{ y - Q[i] - Q , 0 \}.
\end{equation}
Note that $h_i(Q[x]) = g_i(x)$.
If $h_i(y) \leq h_j(y)$ for $y \in \mathbb{R}$, then $j$ is dominated by $i$. \\

\begin{figure}[htbp]
\centering
\includegraphics[width=9cm]{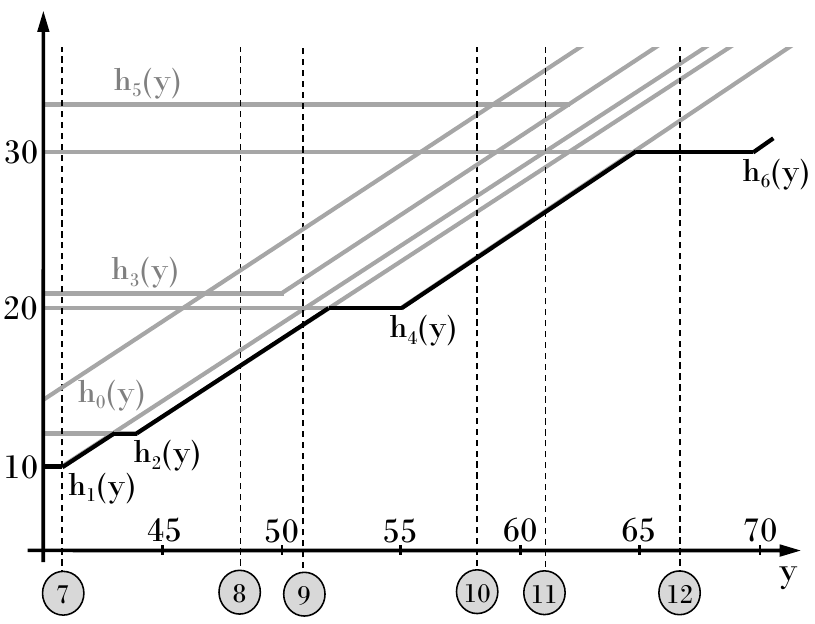}
\caption{Numerical example -- the functions $h_i(y)$ are displayed.}
\label{figure-soft}
\end{figure}

The functions $h_i(y)$ are illustrated in Figure~\ref{figure-soft} for the numerical example of Section~\ref{sectionLinear}.
Each function is piecewise linear and continuous with two pieces: a constant piece with value $h_i = p[i] + d_{0,i+1} - D[i+1]$ for $y \leq Q[i] + Q$ and an increasing piece with slope $\alpha$.

We define two values that characterize the predecessor candidates:
\begin{itemize}[nosep]
\item[]
\begin{itemize}[nosep]
\item the \textbf{fixed cost} $h_i = p[i] + d_{0,i+1} - D[i+1]$, and
\item the \textbf{cumulative demand} $l_i = Q[i]$.
\end{itemize}
\end{itemize}

For $i < j$, $Q[i] \leq Q[j]$ and one can verify that $\{h_i(y) \leq h_j(y) \ \forall y \in \mathbb{R} \} \Leftrightarrow  \{ h_i + \alpha \times (l_j - l_i) \leq h_j \}$.  
For $i > j$, $Q[i] \geq Q[j]$ and one can verify that $\{h_i(y) \leq h_j(y)  \ \forall y \in \mathbb{R} \} \Leftrightarrow \{h_i \leq h_j\}$. These conditions lead to the dominance relation of Equation~(\ref{dominatesFunctionSoft}). We now show that the following loop invariant is respected at line 4 of the algorithm:
\begin{proposition}[Loop Invariant]
$\Lambda$ contains mutually nondominated predecessors of $t$ ranked by increasing fixed cost:
for $i \in \{1,\dots,|\Lambda|-1\}$, $h_{\lambda_i} < h_{\lambda_i+1}$,  $l_{\lambda_i} \leq l_{\lambda_i+1}$, and $h_{\lambda_i} + \alpha (l_{\lambda_j} - l_{\lambda_i}) > h_{\lambda_j}$.  Moreover, $h_{ \lambda_1}(Q[t]) < h_{ \lambda_2}(Q[t])$.
\end{proposition}
This proposition is true for $t = 1$ since $\Lambda = (0)$.
Then, any new node $t$ is inserted at the back of $\Lambda$ only if it is not dominated by the current back element, which implies that $\smash{h_{\lambda_{|\Lambda|}} + \alpha  (l_{\lambda_t} - l_{\lambda_{|\Lambda|}})  > h_{\lambda_t}}$. Then, the algorithm eliminates all dominated nodes $i$ such that $h_{\lambda_t} \leq h_{\lambda_i}$ until it finds the first node that satisfies the invariant condition. Finally, when $t$ is incremented, any front node that does not satisfy the condition $h_{ \lambda_1}(Q[t]) < h_{ \lambda_2}(Q[t])$ is eliminated.

Now we show that this invariant implies that the front node is a best predecessor at each iteration~$t$.
First, $h_{\lambda_1} < h_{\lambda_k}$ and $l_{\lambda_1} \leq l_{\lambda_k}$ for any $k > 1$, so there is no better predecessor in~$\Lambda$.
Second, any other predecessor $i$ that does not appear in $\Lambda$ has either been eliminated because it is dominated by another predecessor in $\Lambda$, or because it was the front element at an iteration $t' \leq t$ and the last condition of Proposition 2 applied. In this specific case, for the second element $j$ we have $ l_i \leq l_j \text{ and } h_j(Q[t']) \leq h_i(Q[t'])$. Because of the shape of the functions $h$, this also implies that  $ h_j(Q[t]) < h_i(Q[t])$ for any $t \geq t'$, so $j$ is an equal or better predecessor. In both cases, the predecessor $i$ has been eliminated from $\Lambda$ only if a better candidate exists, and we have shown that the front element is a best predecessor in $\Lambda$.

\section{Computational Experiments}
\label{sectionExperiments}

The previous section has introduced a linear Split algorithm and its extensions to a limited fleet and soft capacity constraints. 


We now evaluate experimentally the speedup of the new approach compared to the classical Bellman-based algorithm of Section \ref{sectionBellman}.
We generated a set of 105 benchmark instances containing information on the giant tour, the distances between successive nodes, the distances from and to the depot, and finally the demand for each node. Each instance is based on an Euclidean data set from the TSPLib (\url{http://comopt.ifi.uni-heidelberg.de/software/TSPLIB95/}) or the World TSP (\url{http://www.math.uwaterloo.ca/tsp/world/countries.html}). We produced the giant tour using the Lin--Kernighan heuristic of \cite{Helsgaun2000}, and we selected as the depot the node that is the closest to the barycenter of the nodes. We generated the demand for each node randomly with a uniform probability in $[1,50]$.
These instances have between 29 and 71,009 nodes.
For each instance, we considered ten vehicle capacities:\begingroup \setmuskip{\medmuskip}{0mu}
$Q \in \{10^2,2\times10^2, 4\times10^2, 10^3,2\times10^3, 4\times10^3, 10^4,2\times10^4, 4\times10^4,  10^5\}$.
\endgroup
Some (instance, capacity) pairs are eliminated from the set because the capacity of a single vehicle exceeds the total demand of the customers.

We implemented the algorithms of this paper and the original Bellman-based Split in C++.
We implemented the double-ended queue $\Lambda$ as an array of size $n$ with front and back pointers. Overall, these  algorithms use simple data structures and elementary arithmetic, limiting possible bias related to programming style or implementation skills. The code is available at \url{http://w1.cirrelt.ca/\~vidalt/en/VRP-resources.html}. 

We ran the algorithms for each instance and capacity level on a Xeon 3.07\,GHz CPU, using a single thread. The small and medium instances were solved quickly, and so we performed multiple runs in a loop to obtain accurate CPU time measurements. We calibrated the number of runs to achieve a CPU time of about 10 to 60 seconds per instance. \\

\begin{figure}[htbp]
\centering
\includegraphics[width=14.5cm]{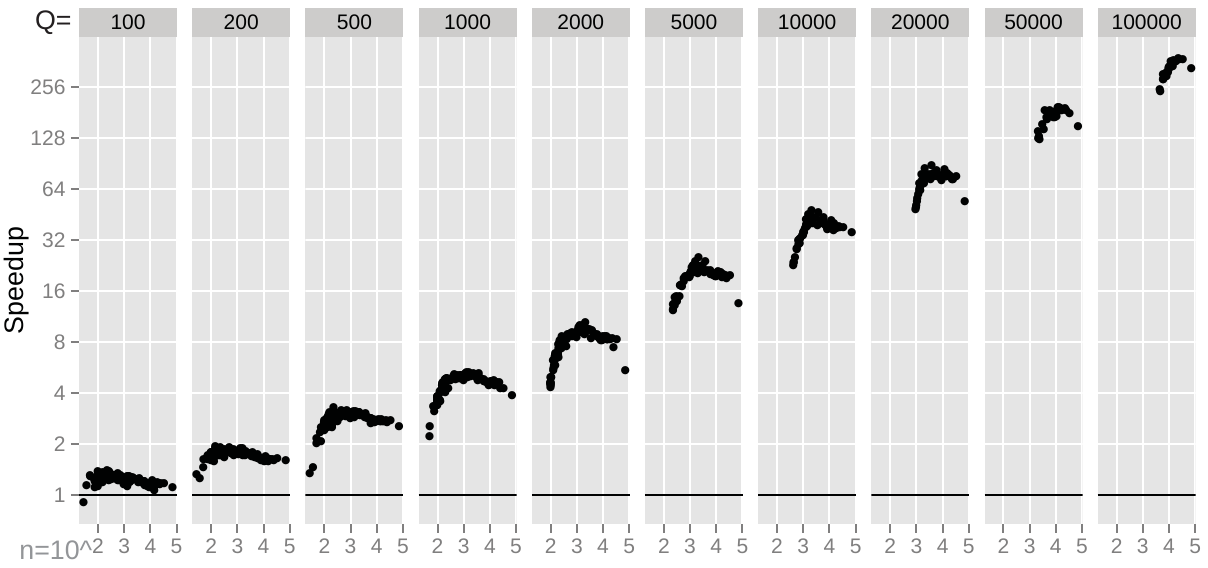}
\caption{Speedups of the linear Split over the Bellman-based algorithm for all 105 instances. The sections of the graph correspond to different values of $Q$. In each section, the X-axis indicates the number of nodes $n$ in each instance. A logarithmic scale~is~used~for~both~axes.}
\label{SpeedupLinear}
\end{figure}

\textbf{Hard capacity constraints.}
Table \ref{Table-Linear} reports the CPU time, in milliseconds, of the Bellman-based Split algorithm and our new linear-time algorithm on a selection of twelve instances. The speedups for all $105 \times 10$ (instance, capacity) pairs are represented graphically in Figure~\ref{SpeedupLinear}, where each section of the figure represents a different capacity $Q$.

\begin{table}[p]
\begingroup
\setmuskip{\medmuskip}{0mu}
\renewcommand{\arraystretch}{1.1}
\caption{CPU time (ms) of Split: using Bellman or linear algorithm} \label{Table-Linear}
\hspace*{-1.45cm}
\scalebox{0.77}
{
\setlength{\tabcolsep}{1.6mm}
\begin{tabular}{|cc|cc|cc|cc|cc|cc|cc|cc}
\multicolumn{12}{c}{\vspace*{-0.5cm}}\\
\hline
\multirow{2}{*}{\textbf{Inst}}&\multirow{2}{*}{\textbf{n}}&\multicolumn{2}{c|}{\textbf{Q=100}}&\multicolumn{2}{c|}{\textbf{Q=200}}&\multicolumn{2}{c|}{\textbf{Q=500}}&\multicolumn{2}{c|}{\textbf{Q=1,000}}&\multicolumn{2}{c|}{\textbf{Q=2,000}}\\
&&\textbf{T$_\textbf{Bellman}$}&\textbf{T$_\textbf{Linear}$}&\textbf{T$_\textbf{Bellman}$}&\textbf{T$_\textbf{Linear}$}&\textbf{T$_\textbf{Bellman}$}&\textbf{T$_\textbf{Linear}$}&\textbf{T$_\textbf{Bellman}$}&\textbf{T$_\textbf{Linear}$}&\textbf{T$_\textbf{Bellman}$}&\textbf{T$_\textbf{Linear}$}\\
\hline
wi29&28&\num{0.00112}&\num{0.00124}&\num{0.00176}&\num{0.00133}&\num{0.0019}&\num{0.001425}&---&---&---&---\\
eil51&50&\num{0.00231}&\num{0.00179}&\num{0.00278}&\num{0.001915}&\num{0.00435}&\num{0.002175}&\num{0.0047}&\num{0.00214}&---&---\\
rd100&99&\num{0.00474}&\num{0.00344}&\num{0.00588}&\num{0.003705}&\num{0.0093}&\num{0.003905}&\num{0.014}&\num{0.00388}&\num{0.018}&\num{0.003965}\\
d198&197&\num{0.00800}&\num{0.00618}&\num{0.01168}&\num{0.00626}&\num{0.0192}&\num{0.00626}&\num{0.0284}&\num{0.00592}&\num{0.0448}&\num{0.00584}\\
fl417&416&\num{0.01552}&\num{0.0124}&\num{0.02312}&\num{0.01304}&\num{0.0404}&\num{0.01402}&\num{0.0704}&\num{0.01428}&\num{0.1152}&\num{0.01406}\\
pr1002&1001&\num{0.0409}&\num{0.03395}&\num{0.0608}&\num{0.03375}&\num{0.0975}&\num{0.0337}&\num{0.173}&\num{0.03505}&\num{0.302}&\num{0.0337}\\
mu1979&1978&\num{0.0843}&\num{0.06835}&\num{0.1174}&\num{0.069}&\num{0.198}&\num{0.0669}&\num{0.341}&\num{0.0678}&\num{0.614}&\num{0.06955}\\
fnl4461&4460&\num{0.1828}&\num{0.15525}&\num{0.2676}&\num{0.1598}&\num{0.454}&\num{0.1618}&\num{0.765}&\num{0.15855}&\num{1.406}&\num{0.15675}\\
kz9976&9975&\num{0.425}&\num{0.3665}&\num{0.614}&\num{0.3745}&\num{1.015}&\num{0.37}&\num{1.62}&\num{0.363}&\num{2.84}&\num{0.3405}\\
d18512&18511&\num{0.799}&\num{0.676}&\num{1.15}&\num{0.714}&\num{1.92}&\num{0.7145}&\num{3.22}&\num{0.6975}&\num{5.84}&\num{0.7065}\\
bm33708&33707&\num{1.519}&\num{1.3015}&\num{2.112}&\num{1.2985}&\num{3.515}&\num{1.2825}&\num{5.48}&\num{1.296}&\num{10.64}&\num{1.295}\\
ch71009&71008&\num{3.39}&\num{3.0615}&\num{4.464}&\num{2.792}&\num{7.54}&\num{2.986}&\num{12.82}&\num{3.339}&\num{22.72}&\num{4.221}\\
\hline
\multicolumn{12}{c}{}\\
\multicolumn{12}{c}{\vspace*{-0.5cm}}\\
\hline
\multirow{2}{*}{\textbf{Inst}}&\multirow{2}{*}{\textbf{n}}&\multicolumn{2}{c|}{\textbf{Q=5,000}}&\multicolumn{2}{c|}{\textbf{Q=10,000}}&\multicolumn{2}{c|}{\textbf{Q=20,000}}&\multicolumn{2}{c|}{\textbf{Q=50,000}}&\multicolumn{2}{c|}{\textbf{Q=100,000}}\\
&&\textbf{T$_\textbf{Bellman}$}&\textbf{T$_\textbf{Linear}$}&\textbf{T$_\textbf{Bellman}$}&\textbf{T$_\textbf{Linear}$}&\textbf{T$_\textbf{Bellman}$}&\textbf{T$_\textbf{Linear}$}&\textbf{T$_\textbf{Bellman}$}&\textbf{T$_\textbf{Linear}$}&\textbf{T$_\textbf{Bellman}$}&\textbf{T$_\textbf{Linear}$}\\
\hline
fl417&416&\num{0.234}&\num{0.01368}&\num{0.316}&\num{0.014}&---&---&---&---&---&---\\
pr1002&1001&\num{0.67}&\num{0.03315}&\num{1.18}&\num{0.0334}&\num{1.74}&\num{0.03425}&---&---&---&---\\
mu1979&1978&\num{1.39}&\num{0.06835}&\num{2.62}&\num{0.066}&\num{4.58}&\num{0.06675}&---&---&---&---\\
fnl4461&4460&\num{3.23}&\num{0.15445}&\num{6.18}&\num{0.14885}&\num{10.84}&\num{0.1378}&\num{24.8}&\num{0.14715}&\num{36}&\num{0.14705}\\
kz9976&9975&\num{7.4}&\num{0.368}&\num{13.1}&\num{0.3435}&\num{27.4}&\num{0.3565}&\num{63.5}&\num{0.3565}&\num{122}&\num{0.367}\\
d18512&18511&\num{13.65}&\num{0.6815}&\num{24.6}&\num{0.6625}&\num{51.6}&\num{0.6775}&\num{125.5}&\num{0.6695}&\num{233}&\num{0.6475}\\
bm33708&33707&\num{24.95}&\num{1.268}&\num{48.6}&\num{1.2825}&\num{95.4}&\num{1.2565}&\num{237.5}&\num{1.342}&\num{464}&\num{1.2485}\\
ch71009&71008&\num{53.05}&\num{3.924}&\num{103}&\num{2.9045}&\num{203.8}&\num{3.8095}&\num{505}&\num{3.4135}&\num{1001}&\num{3.0595}\\
\hline
\multicolumn{12}{c}{\vspace*{0.1cm}}
\end{tabular}
}
\endgroup 
\end{table}

Both algorithms appear to be reasonably fast in the presence of an unlimited fleet and hard capacity constraints. For both methods, the CPU time grows linearly as a function of $n$ when the capacity $Q$ is fixed, since $B$ is also fixed. The time of the  Bellman-based algorithm ranges from a  fraction of milliseconds for small and medium instances with short routes, up to one second for an instance with $71,009$ nodes and $Q=100,000$. For the same instance, the linear split algorithm does not exceed three milliseconds.
Split is used extensively in modern population-based heuristics for the VRP (e.g., between 10,000 and 50,000 times per run in \citealt{Prins2004} and \citealt{Vidal2012}), so a small CPU time is essential.

As illustrated in Figure \ref{SpeedupLinear}, the overall speedup between the linear algorithm and the Bellman-based version grows linearly with the capacity $Q$, which is itself proportional to $B$. The break-even point in terms of route size---beyond which the linear algorithm is faster---is $Q = 100$, when the routes have an average of four customers. Therefore, the linear Split algorithm is beneficial for most VRP applications. For large instances and long routes, the benefits of the proposed Split algorithm are very large, with speedup factors greater than $300$.

Note that the speedup as a function of $n$, for a fixed value of $Q$, is not exactly constant but instead slightly concave. This can be explained by a combination of effects. First, for small values of $n$, the inner loop of the Bellman-based algorithm is slightly faster because it is limited by the end of the giant tour. Second, the CPU time required for initialization and access of the arrays $D[i]$ and $Q[i]$ may not be exactly linear as a function of $n$, due to reduced efficiency of the memory cache on large problems. \\

\begin{figure}[htbp]
\centering
\includegraphics[width=14.5cm]{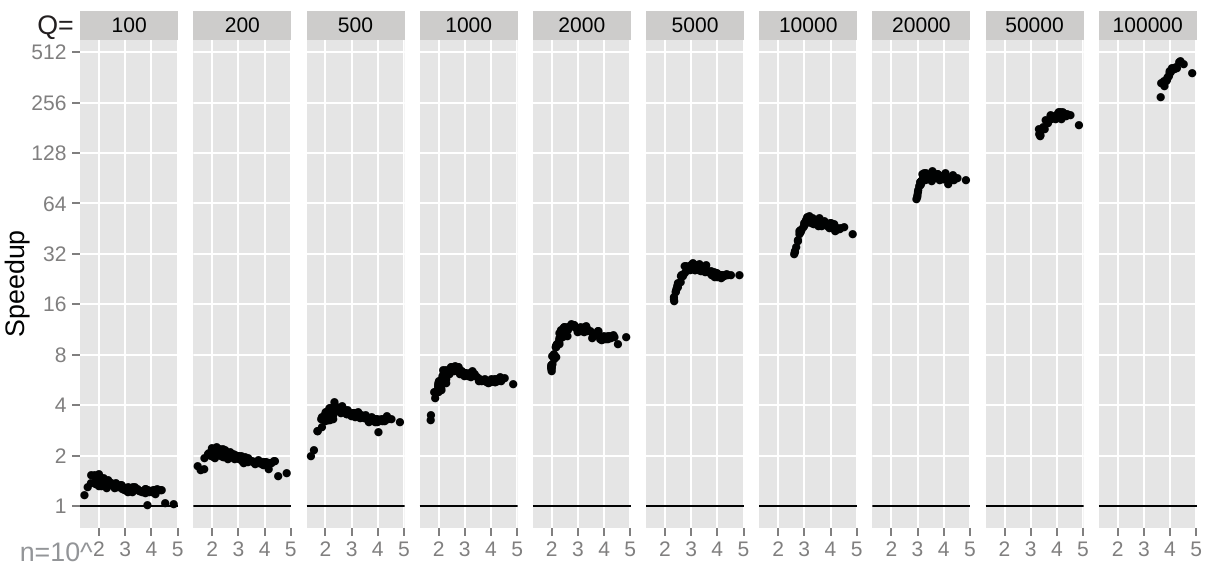}
\caption{Speedup factors for the case with a limited fleet.}
\label{SpeedupBounded}
\end{figure}

\textbf{Limited Fleet.}
Figure \ref{SpeedupBounded} presents results in the same format for the Split algorithm with a limited fleet. In these experiments, the maximum fleet value $m$ is set to the optimal number of vehicles obtained from the unlimited Split algorithm. The algorithm with a limited fleet returns the best solution for any number of routes $k \leq m$.
The conclusions are similar to those of the previous case, with speedup factors ranging from $1$ to $447$. The CPU times are a factor of $m$ higher than in the previous case for all instances.
The largest CPU time for both algorithms, around 58 seconds, occurs for the largest instance with $Q=100$, a regime with short routes but many vehicles, where the Bellman-based and linear Split algorithms perform equally. 
To further reduce the CPU time in these cases, one could rely on advanced algorithms for minimum weight $k$-link paths on graphs with the Monge property \citep{Aggarwal1994}, or explore a heuristic Split based on a Lagrangian relaxation of the fleet-size limit. \\

\pagebreak

\begin{figure}[htbp]
\centering
\vspace*{-0.2cm}
\includegraphics[width=14.5cm]{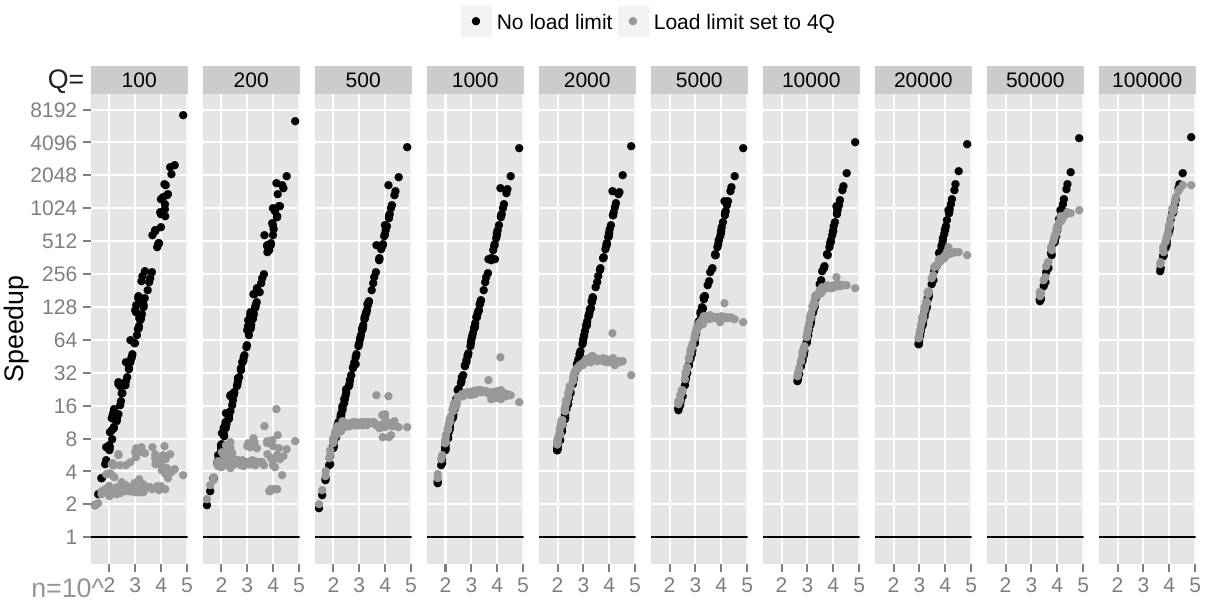}
\caption{Speedups for soft capacity constraints. Two sets of results are presented: the speedups relative to the Bellman algorithm with no limit on the excess capacity (black dots),  and those relative to the Bellman algorithm with a limit of $4Q$ on the total demand of a route (gray dots).}
\label{SpeedupSoft}
\end{figure}

\textbf{Soft capacity constraints.}
Finally, Figure \ref{SpeedupSoft} displays the speedup factors for the Split problem with an unlimited fleet and soft capacity constraints.
\blue{Considering soft capacity constraints puts the Bellman-based algorithm at a larger disadvantage. Indeed, the size of the auxiliary Split graph is not limited anymore by feasibility checks, leading to a complexity of $\Theta(n^2)$ instead of $\Theta(nB)$, while the proposed Split remains linear in all cases. To mitigate this impact, a limit on the capacity excess may be set to reduce the CPU time of the Bellman-based approach. In our experiments, we considered $Q' = 4Q$ and display both sets of results}: the black dots indicate the results of the unlimited case, and the gray dots indicate the results of the limited case.

The $\cO(n^2)$ growth of the Bellman-based Split in the unlimited case can be observed on the figure: it leads to a linear growth of the speedup factor as a function of $n$, up to $7187$ for the largest instance.
The speedups are smaller but still significant when the comparison is with the Bellman algorithm with the $4Q$ bound: from $2$ to $1800$. The maximum CPU time of the linear Split algorithm, for the largest problem instance, is 4.37 milliseconds, compared to 6.3 and 16.6 seconds for the Bellman-based algorithms.

\section{Conclusions}
\label{sectionConclusions}

In this article, we have introduced a simple and efficient Split algorithm in $\cO(n)$.
The algorithm uses dominance properties and can be extended to deal with a limited number of vehicles or relaxed capacity constraints. Our computational experiments show that the new algorithm is significantly faster than the usual Bellman-based approach on VRPs of a realistic size. Positive speedups are encountered when the number of deliveries per route is greater than four. For large problems with 70,000 deliveries and few routes, a speedup factor of up to $400$ is observed.

\blue{There are multiple opportunities for future research.
%
First, one can revisit existing Split-based metaheuristics, measure their new performances, and adapt them to very large-scale CVRP instances.
Several neighborhood-search, neighborhood-pruning and memories techniques \citep{Bentley1992,Toth2003,Irnich2006,Vidal2012b} are known to successfully reduce the complexity of local searches (LS) for large problems. However, the Split algorithm remained, until now, the second most important time bottleneck, and dealing with much larger instances required improvements on both fronts. With the new $\cO(n)$ algorithm, one important barrier has been cleared, and we can focus on further improving the LS.



Second, one can consider more systematic uses of Split in heuristic searches, either by exploring the space of the giant tours \citep{Prins2004,Prins2009c} more intensively, or by using Split as an implicit route-evaluation procedure \citep{Vidal2015,Vidal2014} for VRPs with multiple trips per vehicle, intermediate facilities, or recharging stations.
Similar predecessor-filtering techniques may also be useful for some nonpolynomial versions of Split, e.g., for location-routing problems \citep{Duhamel2010}, VRPs with a heterogeneous fleet \citep{Duhamel2011a} or with decisions on service selections \citep{Vidal2014}.}

Finally, many multi-attribute vehicle routing and scheduling problems, with additional constraints, decision sets, and objectives can be modeled via resources, resource constraints, and their extension functions on the routes \citep{Desaulniers1998a,Irnich2008b}. Based on this formalism, it would be profitable to take a step back and consider current algorithms in a more general perspective, identifying which properties of the extension functions allow for efficient Split algorithms and other neighborhood-evaluation procedures. This is an important task, as the success of modern vehicle routing metaheuristics is, for a large part, conditioned by the computational complexity of their most elementary building blocks.


\end{document}